\documentclass[twocolumn,a4paper]{revtex4}
\usepackage{amsfonts,amsmath,amssymb,mathrsfs,dsfont,yfonts,bbm}
\usepackage[shortlabels]{enumitem}
\usepackage{tikz}
\begin{document}

\newcommand{\Supertwistor}{\Cset \mathrm{P}^{3|4}}
\newcommand{\Twistorspace}{\Cset \mathrm{P}^{3}}
\newcommand{\half}{\frac{1}{2}}
\newcommand{\diff}{\mathrm{d}}
\newcommand{\ra}{\rightarrow}
\newcommand{\Zset}{{\mathbb Z}}
\newcommand{\Cset}{{\,\,{{{^{_{\pmb{\mid}}}}\kern-.47em{\mathrm C}}}}}
\newcommand{\Rset}{{\mathrm{I}\!\mathrm{R}}}
\newcommand{\gra}{\alpha}
\newcommand{\grl}{\lambda}
\newcommand{\gre}{\epsilon}
\newcommand{\zb}{{\bar{z}}}
\newcommand{\mn}{{\mu\nu}}
\newcommand{\Acal}{{\mathcal A}}
\newcommand{\Rcal}{{\mathcal R}}
\newcommand{\Dcal}{{\mathcal D}}
\newcommand{\Mcal}{{\mathcal M}}
\newcommand{\Ncal}{{\mathcal N}}
\newcommand{\Kcal}{{\mathcal K}}
\newcommand{\Lcal}{{\mathcal L}}
\newcommand{\Scal}{{\mathcal S}}
\newcommand{\CW}{{\mathcal W}}
\newcommand{\Bcal}{\mathcal{B}}
\newcommand{\Ccal}{\mathcal{C}}
\newcommand{\Vcal}{\mathcal{V}}
\newcommand{\Ocal}{\mathcal{O}}
\newcommand{\Zcal}{\mathcal{Z}}
\newcommand{\Zb}{\overline{Z}}
\newcommand{\Urm}{{\mathrm U}}
\newcommand{\Srm}{{\mathrm S}}
\newcommand{\SO}{\mathrm{SO}}
\newcommand{\Sp}{\mathrm{Sp}}
\newcommand{\SU}{\mathrm{SU}}
\newcommand{\U}{\mathrm{U}}
\newcommand{\be}{\begin{equation}}
\newcommand{\ee}{\end{equation}}
\newcommand{\Comment}[1]{{}}
\newcommand{\tQ}{\tilde{Q}}
\newcommand{\tq}{{\tilde{q}}}
\newcommand{\trho}{\tilde{\rho}}
\newcommand{\tphi}{\tilde{\phi}}
\newcommand{\Qcal}{\mathcal{Q}}
\newcommand{\tmu}{\tilde{\mu}}
\newcommand{\dbar}{\bar{\partial}}
\newcommand{\p}{\partial}
\newcommand{\eg}{{\it e.g.\;}}
\newcommand{\ie}{{\it i.e.\;}}
\newcommand{\Tr}{\mathrm{Tr}}
\newcommand{\twistor}{\Cset \mathrm{P}^{3}}
\newcommand{\note}[2]{{\footnotesize [{\sc #1}}---{\footnotesize   #2]}}
\newcommand{\CL}{\mathcal{L}}
\newcommand{\CJ}{\mathcal{J}}
\newcommand{\CA}{\mathcal{A}}
\newcommand{\CH}{\mathcal{H}}
\newcommand{\CD}{\mathcal{D}}
\newcommand{\CE}{\mathcal{E}}
\newcommand{\CQ}{\mathcal{Q}}
\newcommand{\CB}{\mathcal{B}}
\newcommand{\CC}{\mathcal{C}}
\newcommand{\CO}{\mathcal{O}}
\newcommand{\CT}{\mathcal{T}}
\newcommand{\CI}{\mathcal{I}}
\newcommand{\CN}{\mathcal{N}}
\newcommand{\CS}{\mathcal{S}}
\newcommand{\CM}{\mathcal{M}}

\parskip 11pt
\title{\Large {\bf Non-Abelian Anyons and Some Cousins of the Arad-Herzog~Conjecture}} 
\author {Matthew Buican, Linfeng Li, and Rajath Radhakrishnan} 
\affiliation{CRST and School of Physics and Astronomy \\ Queen Mary University of London, London E1 4NS, UK\\ }

\begin{abstract}
\noindent
Long ago, Arad and Herzog (AH) conjectured that, in finite simple groups, the product of two conjugacy classes of length greater than one is never a single conjugacy class. We discuss implications of this conjecture for non-abelian anyons in $2+1$-dimensional discrete gauge theories. Thinking in this way also suggests closely related statements about finite simple groups and their associated discrete gauge theories. We prove these statements and provide some physical intuition for their validity. Finally, we explain that the lack of certain dualities in theories with non-abelian finite simple gauge groups provides a non-trivial check of the AH conjecture.
\end{abstract}
\maketitle

\section*{Introduction}
Non-abelian anyons are interesting for a variety of reasons. For example, they naturally appear in quantum field theory descriptions of knot theory \cite{Witten:1988hf}, they are believed to play an important role in the fractional quantum Hall effect \cite{Moore:1991ks}, and they underly a topological form of quantum computation \cite{wang2010topological}. More recently, they have attracted attention as providing possible lessons for quantum gravity \cite{Rudelius:2020orz}.

In this paper we will be exclusively concerned with the appearance of non-abelian anyons in a particular type of $2+1$-dimensional topological quantum field theory (TQFT): discrete gauge theories \cite{Dijkgraaf:1989pz,Roche:1990hs}. These are gauge theories based on some discrete gauge group, $G$, along with a Dijkgraaf-Witten 3-cocycle, $\omega\in H^3(G,U(1))$ (when $\omega$ is cohomologically non-trivial, the theory is said to be \lq\lq twisted"). The basic degrees of freedom are anyonic line operators (i.e., operators supported on one-dimensional loci of spacetime that have non-trivial braiding with each other) of three general types:
\begin{enumerate}
\item {\it Wilson lines}, which carry electric charge labeled by a linear irreducible representation of $G$, $\pi$. These operators have trivial magnetic charge.
\item {\it Magnetic flux lines} carrying magnetic charge labeled by a conjugacy class, $[g]$, of an element $g\in G$ with $g\ne1$. These operators have trivial electric charge. Depending on the choice of $\omega$, such operators may or may not exist.
\item {\it Dyonic lines} (or simply dyons) carrying a magnetic charge labeled by a conjugacy class, $[g]$, of an element $g\in G$ with $g\ne1$ and an electric charge labeled by an, in general, projective representation of the centralizer of $g$, $N_g$. In the case of an untwisted gauge theory (i.e., $\omega=0\in H^3(G,U(1))$), the representation is linear. We will describe, in some detail, when this statement continues to hold for certain dyons in twisted theories. Dyons are the most generic type of anyons in discrete gauge theories.
\end{enumerate}
As a physical toy model, one can think of dyons as Aharonov-Bohm systems with charges bound to magnetic flux lines \cite{Preskill}.

Our first rather basic observation is that the line operators in discrete gauge theories naturally relate close cousins in group theory: representations (and their characters) to centralizers. Therefore, discrete gauge theory is a natural way to organize and unify ideas in the theory of finite groups. 

In what follows, we will focus on the case of finite simple groups. Via group extensions, these are the basic building blocks of all finite groups. The celebrated classification of finite simple groups guarantees that any such group fits into the following categories:
\begin{enumerate}
\item Abelian groups of prime order 
\item Alternating groups
\item Lie groups over finite fields
\item Twenty-six sporadic groups
\end{enumerate}

In spite of this complete classification, there are still many open problems involving these groups. Of particular interest to us is the following old conjecture:

\noindent
{\bf Conjecture (Arad-Herzog):} Consider a non-abelian finite simple group, $G$, and non-trivial elements $g,h\in G$. Then,
\begin{equation}
[g]\cdot [h]\ne [gh]~,
\end{equation}
where $[g]$, $[h]$, and $[gh]$ are conjugacy classes of $g, h,$ and $gh$ respectively \cite{AradHerzog}.

\noindent
More pithily, Arad and Herzog (AH) conjectured that in non-abelian finite simple groups, the product of non-trivial conjugacy classes cannot be a single conjugacy class.

\begin{figure}[h!]

\tikzset{every picture/.style={line width=0.75pt}} 

\begin{tikzpicture}[x=0.75pt,y=0.75pt,yscale=-0.8,xscale=1]

\draw [color={rgb, 255:red, 0; green, 0; blue, 0 }  ,draw opacity=1 ]   (302.88,78.94) -- (331,121.28) ;
\draw [color={rgb, 255:red, 0; green, 0; blue, 0 }  ,draw opacity=1 ]   (359.68,80.44) -- (331,121.28) ;
\draw [color={rgb, 255:red, 0; green, 0; blue, 0 }  ,draw opacity=1 ]   (331,121.28) -- (330.57,168.94) ;

\draw (272,52.4) node [anchor=north west][inner sep=0.75pt]    {$\mathcal{L}_{([ g] ,\pi ^{\omega }_{g})}$};
\draw (344,53.4) node [anchor=north west][inner sep=0.75pt]    {$\mathcal{L}_{([ h] ,\pi ^{\omega }_{h})}$};
\draw (301,173.4) node [anchor=north west][inner sep=0.75pt]    {$\mathcal{L}_{([k] ,\pi ^{\omega }_{k})}$};

\end{tikzpicture}
\caption{Fusion of dyons}
\label{dyonfusion}
\end{figure}
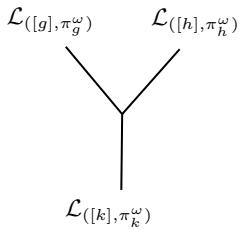

As we will argue in section \ref{conjecture}, this conjecture has the following implication (which we then prove in section \ref{proofs}): 

\noindent
{\bf Theorem 1:} In a (twisted or untwisted) $2+1$-dimensional discrete gauge theory with a non-abelian finite simple gauge group, the fusion of any two lines carrying non-trivial magnetic flux (as in Figure \ref{dyonfusion}) cannot have a unique fusion outcome.

In other words, Theorem 1 asserts we cannot have
\begin{equation}\label{fusionL}
\CL_{([g],\pi^{\omega}_g)}\times \CL_{([h],\pi_h^{\omega})}=\CL_{([k],\pi_{k}^{\omega})}~,\ \ \ g,h\ne1~,
\end{equation}
where, generically, all lines (denoted by $\CL$) are non-abelian dyons \footnote{Of course, we may allow for pure fluxes to appear in \eqref{fusionL}.}. We will think of this theorem as a first cousin of the AH conjecture.

\begin{figure}[h!]

\tikzset{every picture/.style={line width=0.75pt}} 

\begin{tikzpicture}[x=0.75pt,y=0.75pt,yscale=-0.8,xscale=1]

\draw [color={rgb, 255:red, 45; green, 166; blue, 175 }  ,draw opacity=1 ]   (308.76,81.69) -- (336.89,124.03) ;
\draw [color={rgb, 255:red, 45; green, 166; blue, 175 }  ,draw opacity=1 ]   (365.57,83.19) -- (336.89,124.03) ;
\draw [color={rgb, 255:red, 45; green, 166; blue, 175 }  ,draw opacity=1 ]   (336.89,124.03) -- (336.46,171.69) ;

\draw (295.94,58.85) node [anchor=north west][inner sep=0.75pt]    {$\mathcal{W}_{\pi }$};
\draw (322.94,175.85) node [anchor=north west][inner sep=0.75pt]    {$\mathcal{W}_{\pi''}$};
\draw (354.94,59.85) node [anchor=north west][inner sep=0.75pt]    {$\mathcal{W}_{\pi ^{'}}$};

\end{tikzpicture}

\caption{Fusion of Wilson lines}
\label{Wilsonfusion}
\end{figure}
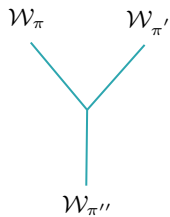

So far, we have avoided discussing the fusion of Wilson lines. However, in light of \eqref{fusionL}, it is interesting to ask if we can fuse non-abelian Wilson lines $\CW_{\pi}$ and $\CW_{\pi'}$ (as in Figure \ref{Wilsonfusion}) to obtain a unique outcome
\begin{equation}\label{abcW}
\CW_{\pi}\times\CW_{\pi'}=\CW_{\pi''}~.
\end{equation}

As we will briefly explain in section \ref{fusionM}, \eqref{abcW} is equivalent to demanding that, at the level of group theory
\begin{equation}\label{charF}
\chi_{\pi}\cdot\chi_{\pi'}=\chi_{\pi''}~,
\end{equation}
where $\chi_{\pi},\chi_{\pi'}$, and $\chi_{\pi''}$ are, respectively, the characters of irreducible linear representations, $\pi, \pi'$, and $\pi''$, of $G$ with dimension greater than 1. Although it might seem strange that \eqref{charF} is possible (especially if one thinks of taking products of irreducible representations in $SU(N)$), it turns out that products of irreducible representations of finite simple groups can be irreducible \cite{zisser1993irreducible}.

The corresponding (twisted or untwisted) discrete gauge theory then has a product of Wilson lines as in \eqref{abcW}. One simple example of this phenomenon in theories with a non-abelian simple gauge group involves the fusion of a Wilson line carrying charge in the $8$-dimensional representation of $A_9$ with a Wilson line carrying charge in either of the $21$-dimensional representations. Intriguingly, the discrete gauge theories based on finite simple groups are prime \cite{naidu2009fusion}, so they do not consist of separate TQFTs with trivial mutual braiding. Therefore, \eqref{abcW} corresponds to some other structural properties of the $A_9$ discrete gauge theory. We will discuss these properties more generally in an upcoming work \cite{toAppear}. 

\begin{figure}[h!]

\tikzset{every picture/.style={line width=0.75pt}} 

\begin{tikzpicture}[x=0.75pt,y=0.75pt,yscale=-0.8,xscale=1]

\draw [color={rgb, 255:red, 45; green, 166; blue, 175 }  ,draw opacity=1 ]   (298.7,90.44) -- (326.82,132.78) ;
\draw [color={rgb, 255:red, 208; green, 2; blue, 27 }  ,draw opacity=1 ]   (355.5,91.94) -- (326.82,132.78) ;
\draw [color={rgb, 255:red, 0; green, 0; blue, 0 }  ,draw opacity=1 ]   (326.82,132.78) -- (326.39,180.44) ;

\draw (283.94,68.85) node [anchor=north west][inner sep=0.75pt]    {$\mathcal{W}_{\pi }$};
\draw (345,70.4) node [anchor=north west][inner sep=0.75pt]    {$\mu _{[ g]}$};
\draw (300,183.4) node [anchor=north west][inner sep=0.75pt]    {$\mathcal{L}_{([ h] ,\pi _{h}^{\omega})}$};

\end{tikzpicture}

\caption{Fusion of a Wilson line with a magnetic flux line}
\label{Wilsonfluxfusion}
\end{figure}
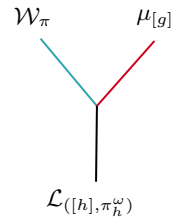

Therefore, we learn that a version of the AH conjecture for characters alone cannot hold. However, our physical discussion above suggests studying one more type of fusion with a unique outcome (Figure \ref{Wilsonfluxfusion})
\begin{equation}\label{abcWd}
\CW_{\pi}\times\CL_{([g],\pi_g^{\omega})}=\CL_{([h],\pi^{\omega}_h)}~, \ \ \ g\ne1~,
\end{equation}
where $\CW_{\pi}$ is a non-abelian Wilson line, and the remaining anyons are non-abelian dyons. As a slightly simpler fusion, we may study the following fusion with a unique outcome
\begin{equation}\label{abcWmu}
\CW_{\pi}\times\mu_{[g]}=\CL_{([h],\pi_h^{\omega})}~, \ \ \ g\ne1~,
\end{equation}
where we have replaced the dyon on the left-hand side of \eqref{abcWd} with a non-abelian flux line. Here we have implicitly assumed that the flux line also exists in the theory (depending on the twist, this assumption may or may not hold).

This observation brings us to our second cousin of the AH conjecture:

\noindent
{\bf Theorem 2:} In any (twisted or untwisted) discrete gauge theory based on a non-abelian finite simple group, $G$, fusion of the types in \eqref{abcWd} and \eqref{abcWmu} is forbidden.

\noindent
{\bf Intuition:} One heuristic intuition behind this theorem is the following. As a consequence of theorem 1, theorem 2 implies that in discrete gauge theories based on non-abelian simple groups, the only allowed fusions with unique outcomes involving non-abelian anyons are those in \eqref{abcW}. Wilson lines have trivial braiding amongst themselves \footnote{Physically, this last statement is clear from the fact that Wilson lines do not carry magnetic flux. In the language of category theory, this statement follows from the well-known fact that Wilson lines form a symmetric fusion subcategory. In fact, it is a Lagrangian subcategory isomorphic to the category of finite dimensional representations of $G$ over $\mathbb{C}$, ${\rm Rep}(G)$.}. Therefore, even though the fusion in \eqref{abcW} does not arise from a factorization of the TQFT into separate theories with trivial mutual braiding, the Wilson lines themselves have trivial mutual braiding.

Just as theorem 1 follows from the AH conjecture, so too theorem 2 follows from a more basic theorem on finite simple groups which we refer to as the third cousin of the AH conjecture:

\noindent
{\bf Theorem 3:} Consider any non-abelian finite simple group, $G$, any irreducible linear representation, $\pi$, of $G$ having dimension greater than one, and the centralizer, $N_g$, of any $g\ne1$. The restricted representation, $\pi|_{N_g}$, is reducible.

\noindent
We refer to theorems 1-3 as \lq\lq cousins" of the AH conjecture since they are all related by TQFT. 

Note that the above discussion is not relevant for abelian simple groups (the type 1 finite simple groups in the above-described classification) since these groups do not have conjugacy classes of length larger than one or representations of dimension larger than one. In other words, their fusion rules are those of a discrete finite group. As a result, we focus on non-abelian finite simple groups.

\noindent
{\bf Duality:} It is also interesting to understand how our above picture is compatible with a type of electric/magnetic duality that often features in discrete gauge theories. For example, the $S_3$ discrete gauge theory has a duality that exchanges the Wilson line charged under the 2-dimensional representation with the line having flux in the 3-cycle conjugacy class \cite{beigi2011quantum,nikshych2014categorical}. More general examples have been discussed in \cite{nikshych2014categorical,naidu2007categorical,Hu:2020ghf}. Clearly, theorems 1, 2, and 3 can only be compatible with such dualities if the Wilson lines participating in \eqref{abcW} are not exchanged with lines carying non-abelian flux. In fact, no such dualities exist in theories based on non-abelian finite simple gauge groups ({\bf Proof:} apply theorem 5.8 of \cite{naidu2007categorical} noting that non-abelian simple groups have no non-trivial abelian normal subgroups). This fact is a non-trivial check of the above picture and is a check of the AH conjecture (this latter claim holds since, if theorem 1 were not true, then the AH conjecture would be false) \footnote{This discussion does not preclude a duality between a twisted discrete gauge theory with a finite simple gauge group and some other type of TQFT (although presumably the dualities should live inside the class of theories considered in \cite{Freed:2020qfy}).}.

The structure of this paper is as follows. In the next section, we give a physicist's introduction to the machinery behind the fusion rules that are relevant for proving the above theorems and deriving theorem 1 from the AH conjecture. In section \ref{fusionM}, we give a brief overview for the more mathematically inclined reader. We derive theorem 1 from the AH conjecture in section \ref{conjecture} before proceeding to proofs of theorems 1, 2, and 3 in section \ref{proofs}. Theorem 3 turns out to be equivalent to theorem 2 and has its own, purely group theoretical proof, that we describe in section \ref{proofs}. The appendices include various additional checks we performed on theorem 3 and therefore 2.

\section{Physics: fusion in discrete gauge theories}\label{fusionP}
One modern perspective on how to go from a group, $G$, to a $2+1$-dimensional discrete gauge theory is to start from a $G$-symmetry-protected topological phase ($G$-SPT) and gauge $G$ \cite{Barkeshli:2014cna}. At the same time, it may be useful to keep in mind that many of the results we will need in this section predate this perspective and follow from the classic work \cite{Dijkgraaf:1989pz}.

The starting point is a set of surface defects in one-to-one correspondence with the elements $g\in G$. For simplicity, we label these defects by group elements as well. Fusion of these defects obeys the usual group multiplication law of $G$, so $g\times h=gh$. One may also consider deforming the associativity of defect fusion via a 3-cocycle
\begin{equation}\label{3cocycle}
\omega(g,h,k)\in H^3(G,U(1))~.
\end{equation}
The $H^3(G,U(1))$ cohomology group then labels the distinct $2+1$-dimensional $G$-SPTs.

Gauging $G$ corresponds to constructing conjugacy classes, $[g_i]$, for a set of representative $g_i\in G$ and pairing this data with an irreducible representation, $\pi^{\omega}_{g_i}$, of the centralizer of each $g_i$, $N_{g_i}$. These are, respectively, the magnetic and electric charges of the discrete gauge theory. The 3-cocycle in \eqref{3cocycle} is the Dijkgraaf-Witten twist (when $\omega=0$ in cohomology we have an untwisted gauge theory).

In this way, lines bounding the $G$-SPT surface operators are liberated and become anyons in the---depending on $\omega$---twisted or untwisted $G$ discrete gauge theory. These latter objects are given by the pair $([g], \pi^{\omega}_{g})$, where the square brackets around $g$ are there to emphasize that we are dealing with a conjugacy class (for any representative in $[g]$, the corresponding centralizers are isomorphic).

The question of whether the electric charge, $\pi^{\omega}_g$, is projective is determined by the reduction of $\omega$ to $N_g$
\be
\label{2cocyclefrom3}
\eta_g(h,k):=\frac{\omega(g,h,k)\omega(h,k,g)}{\omega(h,g,k)}\in H^2(N_g,U(1))~,
\ee
where $h,k\in N_g$. Indeed, this is the phase that appears in
\begin{equation}\label{etarep}
\pi^{\omega}_g(h)\pi^{\omega}_g(k)=\eta_g(h,k)\pi^{\omega}_g(hk)~.
\end{equation}
If $\eta_g$ is trivial in $H^2(N_g,U(1))$ the representation is linear \footnote{If $\eta_g(h,k)$ is a non-trivial 2-coboundary, the projective representations obtained will be in one-to-one correspondence with linear representations. These projective factors can be removed using a symmetry gauge transformation as detailed in \cite{Barkeshli:2014cna}. On the other hand, if $\eta_g(h,k)$ is non-trivial in cohomology, then the corresponding representations of $N_g$ must be higher dimensional.  To see this statement, suppose this were not the case. Then, solving \eqref{etarep} for $\eta_{g}(h,k)$ implies that $\eta_g$ is a 2-coboundary.}. For example, the group $PSL(2,4)$ has $\mathbb{Z}_3$ as the centralizer of elements in its length twenty conjugacy class. Since $H^2(\mathbb{Z}_3,U(1))=\mathbb{Z}_1$, the corresponding $\eta_g$ is trivial regardless of the choice of $\omega\in H^3(PSL(2,4),U(1))\simeq\mathbb{Z}_6\times\mathbb{Z}_{10}$. More generally, if $\omega$ is cohomologically non-trivial, then $\pi^{\omega}_g$ is typically projective.

In light of the discussion in the introduction, the most important thing for us to understand is the fusion of two anyons, $([g],\pi^{\omega}_g)$ and $([h],\pi^{\omega}_h)$. Intuitively, we have to fuse both the conjugacy classes as well as the representations that the anyons depend on. This involves identifying the conjugacy classes of the elements obtained by multiplying the elements in $[g]$ and $[h]$. Also, we have to consistently decompose the product $\pi^{\omega}_g \otimes \pi^{\omega}_h$ into irreducible representations of centralizers of $G$. The precise way to carry out these steps is given by \cite{Barkeshli:2014cna,Dijkgraaf:1989pz}
\begin{eqnarray}\label{genfusion}
N_{([g],\pi^{\omega}_g),([h], \pi^{\omega}_h)}^{([k],\pi^{\omega}_k)}&=&\sum_{(t,s)\in N_g \backslash G /N_h} m(\pi^{\omega}_k |_{N_{{}^tg} \cap N_{{}^sh} \cap N_k},\cr && {}^t\pi^{\omega}_{g}|_{N_{{}^tg} \cap N_{{}^sh} \cap N_k} \otimes {}^s\pi^{\omega}_{h}|_{N_{{}^tg} \cap N_{{}^sh} \cap N_k} \cr&&\otimes \pi^{\omega}_{({}^tg,{}^sh,k)})~,
\end{eqnarray}
where ${}^t\pi^{\omega}_{g}|_{N_{{}^tg} \cap N_{{}^sh} \cap N_k} \otimes {}^s\pi^{\omega}_{h}|_{N_{{}^tg} \cap N_{{}^sh} \cap N_k} \otimes \pi^{\omega}_{({}^tg,{}^sh,k)}$ and $\pi^{\omega}_k |_{N_{{}^tg} \cap N_{{}^sh} \cap N_k}$ are (in general reducible) representations of $N_{{}^tg} \cap N_{{}^sh} \cap N_k$ (${}^t\pi^{\omega}_g$, ${}^s\pi^{\omega}_h$, and $\pi^{\omega}_k$ are representations of $N_{{}^tg}$, $N_{{}^sh}$, and $N_k$ which are then restricted to the intersection subgroup). Here we define ${}^tg:=t^{-1}gt$. The projectivity of the ${}^t\pi^{\omega}_{g}$, ${}^s\pi^{\omega}_{h}$, and $\pi^{\omega}_k$ representations is determined by the corresponding cohomology as in \eqref{2cocyclefrom3}. The representation $\pi^{\omega}_{({}^tg,{}^sh,k)}$ is one dimensional (it is a representation of the action of symmetries on the one-dimensional $V_{{}^tg{}^sh}^{k}$ fusion space in the $G$-SPT) and satisfies
\begin{eqnarray}
\pi^{\omega}_{({}^tg,{}^sh,k)}(\ell)\pi^{\omega}_{({}^tg,{}^sh,k)}(m)&=&{\eta_{k}(\ell,m)\over\eta_{{}^tg}(\ell,m)\eta_{{}^sh}(\ell,m)}\cr && \hspace{0.15cm} \cdot ~ \pi^{\omega}_{({}^tg,{}^sh,k)}(\ell m)~.
\end{eqnarray}
These projective factors guarantee that the two arguments of the $m(\cdot,\cdot)$ function can be meaningfully compared. Roughly speaking, this $m(\cdot,\cdot)$ function computes the inner products of the representations appearing as arguments (see \cite{Barkeshli:2014cna} for further details). Finally, let us note that the sum in \eqref{genfusion} is over the double coset, $N_g\backslash G/N_h$. 

Another closely related quantity of interest is the modular data of a (twisted or untwisted) discrete gauge theory. It is given by \cite{Hu:2012wx}
\begin{eqnarray}
\label{smatrix}
S_{([g],\pi_{g}^{\omega}),([h],\pi_{h}^{\omega})}&=&\frac{1}{|G|}\sum_{\substack{k \in [g],\ \ell \in [h],\\ k \ell=\ell k}} \chi^{k}_{\pi_g^{\omega}}(\ell)^{*} \chi^{\ell}_{\pi_{h}^{\omega}}(k)^{*}~~, \nonumber\\ \theta_{([g], \pi_{g}^{\omega})}&=&\frac{\chi_{\pi^{\omega}_{g}}(g)}{\chi_{\pi^{\omega}_{g}}(e)}~,
\end{eqnarray}
where $\chi^{h}_{\pi_g^{\omega}}(\ell)$ is defined through the relation 
\begin{equation}
\label{projchar}
\chi^{xgx^{-1}}_{\pi_g^{\omega}}(xhx^{-1}):= \frac{\eta_g(x^{-1},xhx^{-1})}{\eta_g(h,x^{-1})} \chi_{\pi^{\omega}_g}(h)~.
\end{equation}
 Here, $\theta$ is the topological spin, and $S$ is the modular $S$ matrix. From these definitions, one can check that the quantum dimensions of the anyons are
\begin{equation}
\label{qdim}
d_{([g],\pi^{\omega}_g)}={S_{([g],\pi_g^{\omega})([1],1)}\over S_{([1],1)([1],1)}}=|[g]|\cdot\text{deg }\pi^{\omega}_g~,
\end{equation}
where $|[g]|$ is the size of $[g]$, and $|\pi^{\omega}_g|$ is the dimension of $\pi^{\omega}_g$. Non-abelian anyons have $d_{([g],\pi^{\omega})}>1$ and necessarily satisfy
\begin{equation}
([g],\pi^{\omega}_g)\times([g^{-1}],(\pi^{\omega}_g)^*)=([1],1)+\cdots~,
\end{equation}
where the ellipses necessarily contain additional terms, $1$ is the trivial representation of $G$, and $(([g^{-1}],(\pi^{\omega}_g)^*)$ is the anyon conjugate to $([g],\pi^{\omega}_g)$.

As we will see in more detail when we prove theorems 1 and 2, anyons $([g],\pi_g^{\omega})$ and $([h],\pi_h^{\omega})$ that fuse to give a unique outcome satisfy the following condition with respect to the $S$ matrix
\be
\label{Smatrixcond}
|S_{([g]\pi_g^{\omega}),([h],\pi_h^{\omega})}|=\frac{1}{|G|} d_{([g],\pi_g^{\omega})}d_{([h],\pi_h^{\omega})}~.
\ee
Let us explore the consequences of this relation. To that end, using \eqref{qdim}, we have $d_{([g],\pi_g^{\omega})}d_{([h],\pi_h^{\omega})}=|[g]||[h]|\cdot\text{deg }\pi^{\omega}_g \cdot\text{deg }\pi^{\omega}_h$. Substituting in \eqref{Smatrixcond} and using \eqref{smatrix}, we have
\begin{eqnarray}
&\frac{1}{|G|}& |[g]||[h]|\cdot\text{deg }\pi^{\omega}_g \cdot\text{deg }\pi^{\omega}_h \nonumber \\
&=&\bigg |\frac{1}{|G|}\sum_{\substack{k \in [g],\ \ell \in [h],\\ k \ell=\ell k}} \chi^{k}_{\pi_g^{\omega}}(\ell)^{*} \chi^{\ell}_{\pi_{h}^{\omega}}(k)^{*} \bigg | \nonumber\\
&\leq & \frac{1}{|G|}\sum_{\substack{k \in [g],\ \ell \in [h],\\ k \ell=\ell k}} |\chi^{k}_{\pi_g^{\omega}}(\ell)| |\chi^{\ell}_{\pi_{h}^{\omega}}(k)|\nonumber \\
&\leq& \frac{|[g]||[h]|}{|G|}  \cdot\text{deg }\pi^{\omega}_g \cdot\text{deg }\pi^{\omega}_h
\end{eqnarray}
In the last inequality above, we have used \eqref{projchar} as well as the fact that projective characters satisfy $|\chi_{\pi^{\omega}_g}|\leq \text{deg } \pi_{g}^{\omega}$ \footnote{This statement is guaranteed as long as the projection factors defining the representations are roots of unity, which is satisfied in our case. Indeed, the 3-cocycle $\omega \in H^3(G,U(1))$ can be chosen to be valued in roots of unity without loss of generality.}. It is clear that \eqref{Smatrixcond} is satisfied if and only if the conjugacy classes $[g]$ and $[h]$ commute element-wise and the projective characters satisfy
\be
\label{charconst}
|\chi_{\pi^{\omega}_g}(l)|= \text{deg } \pi_{g}^{\omega} \text{ and } |\chi_{\pi^{\omega}_h}(k)|= \text{deg } \pi_{h}^{\omega} ~ 
\ee 
$\forall ~ l\in [h], k \in [g]$. This result is a generalization of lemma 3.4 of \cite{naidu2009fusion}. As mentioned above, it will be crucial for our proofs of theorems 1 and 2.

In the language used in this section, we have that non-abelian Wilson lines, flux lines, and dyons correspond to
\begin{eqnarray}\label{dictionary}
\CW_{\pi_1}&\leftrightarrow&([1],\pi_1)~, \ \ \ |\pi_1|>1~,\cr
\mu_{[g]}&\leftrightarrow&([g],1_g^{\epsilon})~, \ \ \ |[g]|>1~,\cr
\CL_{([h],\pi_h^{\omega})}&\leftrightarrow&([h],\pi_h^{\omega})~,\ \ \ |[h]|\cdot|\pi_h^{\omega}|>1~.
\end{eqnarray}
We have dropped the $\omega$ superscript from $\pi_1$ in order to emphasize that the Wilson lines always transform under linear representations of $G$.
We attach the $\epsilon$ superscript on the trivial representation of the flux line because these objects only exist when the relevant $\eta_g$ in \eqref{2cocyclefrom3} is trivial in cohomology, and hence of the form $\eta_g(h,k)= \frac{\epsilon_g(h)\epsilon_g(k)}{\epsilon_g(h\cdot k)}$. Finally, $1_g^{\epsilon}$ is the irreducible projective representation of $N_g$ whose character is proportional to the trivial representation of $N_g$.

As a final comment, we note that, from the above modular data, it is easy to show that 
\begin{equation}
\theta_{\CW_{\pi}}=1~, \ \ \ {S_{\CW_{\pi}\CW_{\pi'}}\over S_{\CW_1\CW_{\pi'}}}=1~,
\end{equation}
where $\CW_1=([1],1)$ is the trivial Wilson line. In other words, as alluded to in the introduction, the Wilson lines are all bosons and have trivial mutual braiding with each other (they have non-trivial braiding with other lines in the theory).

\section{Math: fusion in $\mathcal{Z}({\rm Vec}_{\omega}^G)$}\label{fusionM}
As a guide to the more mathematically inclined reader, we summarize certain aspects of the previous section in a more formal way. The reader who is only interested in the results or a more physical perspective on them is free to skip this section.

The mathematical framework underlying the physical discussion of \cite{Barkeshli:2014cna} used in the last section is the notion of a $G$-crossed braided category \cite{etingof2009fusion}. The gauging procedure corresponds to the mathematical notion of equivariantization.

In fact, to construct the (twisted or untwisted) discrete gauge theory we can use a simpler notion. Our starting point is the category of $G$-graded vector spaces, ${\rm Vec}^G_{\omega}$, with associator given by the $\mathbb{C}^{\times}$-valued 3-cocycle, $\omega$. The theory we obtain upon equivariantization is the modular tensor category (MTC) constructed by the process of taking the Drinfeld center \cite{bakalov2001lectures,ostrik2002module}. In particular, our gauge theory is just
\begin{equation}\nonumber
{\rm Twisted}\ G\ {\rm discrete\ gauge\ theory}\leftrightarrow\mathcal{Z}({\rm Vec}^G_{\omega})~.
\end{equation}
The various operators discussed in \eqref{dictionary} correspond to the simple objects of $\mathcal{Z}({\rm Vec}^G_{\omega})$ with categorical dimension larger than one. The simple objects corresponding to the trivial conjugacy class in $G$ (what we have called Wilson lines) have trivial topological spin, $\theta$, and are closed under fusion. This means they form a symmetric subcategory. In fact, as is well-known, these simple objects form a Lagrangian subcategory isomorphic to ${\rm Rep}(G)$, the category of finite dimensional representations of $G$ over $\mathbb{C}$. In particular, Wilson line fusion rules are those of the corresponding representation semiring. This observation explains the equivalence of \eqref{abcW} and \eqref{charF}.

\section{From fusion to theorem 1 and a relation between theorems 2 and 3}\label{conjecture}
Given the construction in section \ref{fusionP}, we will first explain why the AH conjecture implies that, in (twisted and untwisted) discrete gauge theories based on simple groups, the fusion of any two lines carrying magnetic flux must have more than one fusion outcome (i.e., theorem 1). After explaining this fact, we will explain the relation between theorems 2 and 3.

To understand the connection between (twisted and untwisted) discrete gauge theories and the AH conjecture, recall the fusion formula in \eqref{genfusion}. Since the arguments of the $m(\cdot,\cdot)$ function are representations of $N_{{}^tg} \cap N_{{}^sh} \cap N_k$, we can decompose them in terms of irreducible representations, $\pi^{\omega(i)}$, of this group
\begin{eqnarray}
&&{}^t\pi^{\omega}_{g}|_{N_{{}^tg} \cap N_{{}^sh} \cap N_k} \otimes {}^s\pi^{\omega}_{h}|_{N_{{}^tg} \cap N_{{}^sh} \cap N_k} \otimes \pi^{\omega}_{({}^tg,{}^sh,k)} \cr
&& \hspace{0.08cm} =\sum_i \alpha_i \pi^{\omega(i)} ~,\cr
&&\pi^{\omega}_k |_{N_{{}^tg} \cap N_{{}^sh} \cap N_k}=\sum_i \alpha^{'}_i \pi^{\omega(i)}~,
\end{eqnarray}
for some non-negative integers $\alpha_i, \alpha^{'}_i$. Then the definition of $m(\cdot,\cdot)$ in \cite{Barkeshli:2014cna} implies
\begin{eqnarray}
 &&m(\pi^{\omega}_k |_{N_{{}^tg} \cap N_{{}^sh} \cap N_k}, {}^t\pi^{\omega}_{g}|_{N_{{}^tg} \cap N_{{}^sh} \cap N_k} \otimes {}^s\pi^{\omega}_{h}|_{N_{{}^tg} \cap N_{{}^sh} \cap N_k} \cr && \hspace{0.05cm} \otimes ~ \pi^{\omega}_{({}^tg,{}^sh,k)})= \sum_i \alpha_i \alpha^{'}_i~.
\end{eqnarray}
We know that $\pi^{\omega}_k$ is an irreducible representation of $N_k$. Also, $N_{{}^tg} \cap N_{{}^sh} \cap N_k$ is a subgroup of $N_k$. According to the Frobenius reciprocity theorem for projective representations of finite groups \cite{gregory} \footnote{We use this theorem in the twisted case; in the untwisted case we use the usual theorem for linear representations.}, we know that, given any irreducible representation, $\pi^{\omega(i)}$, of $N_{{}^tg} \cap N_{{}^sh} \cap N_k$, there is always an irreducible representation, $\pi^{\omega}_k$, of $N_k$ such that the decomposition of $\pi^{\omega}_k |_{N_{{}^tg} \cap N_{{}^sh} \cap N_k}$ into irreducible representations of $N_{{}^tg} \cap N_{{}^sh} \cap N_k$ contains $\pi^{\omega(i)}$. This reasoning shows that, given ${}^t\pi^{\omega}_{g}|_{N_{{}^tg} \cap N_{{}^sh} \cap N_k} \otimes {}^s\pi^{\omega}_{h}|_{N_{{}^tg} \cap N_{{}^sh} \cap N_k} \otimes \pi^{\omega}_{({}^tg,{}^sh,k)}$, there is always some irreducible representation, $\pi^{\omega}_k$, such that $ m(\pi^{\omega}_k |_{N_{{}^tg} \cap N_{{}^sh} \cap N_k}, {}^t\pi^{\omega}_{g}|_{N_{{}^tg} \cap N_{{}^sh} \cap N_k} \otimes {}^s\pi^{\omega}_{h}|_{N_{{}^tg} \cap N_{{}^sh} \cap N_k} \otimes \pi^{\omega}_{({}^tg,{}^sh,k)})$ is non-zero. It follows that once we choose some conjugacy class, $[k]$, such that $[k] \in [g] \cdot [h]$, there is always some $\pi^{\omega}_k$ such that $N_{([g],\pi^{\omega}_g)([h],\pi^{\omega}_h)}^{([k],\pi^{\omega}_k)}\neq 0$. Here, $[g] \cdot[h]$ are the conjugacy classes obtained from taking a product of anyons with magnetic charges in $[g]$ and $[h]$. 

Hence, in order to have a fusion rule of the type
\be
\label{fusgen}
([g], \pi^{\omega}_g) \times ([h], \pi^{\omega}_h)= ([k], \pi^{\omega}_k)~,\ \ \ g,h\ne1~,
\ee
where all magnetic fluxes on the LHS are non-trivial, we need the fusion of the orbits $[g] \cdot [h]$ to contain only a single orbit $[k]$ (note that $|[k]|$ need not be equal to $|[g]||[h]|$ \footnote{In the case of the fusion of pure fluxes, we do require $|[k]|=|[g]||[h]|$.}). Moreover, commutativity of the fusion rules requires $[k]=[h]\cdot [g]$. Hence, the double coset $N_g \backslash G /N_h$ should have only a single element. (Since the double coset is trivial, we will remove the $t,s$ dependence in the expressions below). 
We also require that the decomposition of representations $\pi^{\omega}_k |_{N_g \cap N_h \cap N_k}$ and $\pi^{\omega}_{g}|_{N_g \cap N_h \cap N_k} \otimes \pi^{\omega}_{h}|_{N_g \cap N_h \cap N_k} \otimes \pi^{\omega}_{(g,h,k)}$ into irreps of $N_g \cap N_h \cap N_k$ to have only a single irrep (of multiplicity one) in common. That is, if 
\begin{eqnarray}
\pi^{\omega}_{g}|_{N_{g} \cap N_{h} \cap N_k} \otimes \pi^{\omega}_{h}|_{N_{g} \cap N_{h} \cap N_k} \otimes \pi^{\omega}_{(g,h,k)}= \sum_i \alpha_i \pi^{\omega(i)} \nonumber\ \ \ \ \\
\pi^{\omega}_k |_{N_{g} \cap N_{h} \cap N_k}= \sum_i \alpha^{'}_i \pi^{\omega(i)}~,\ \ 
\end{eqnarray}
then there should be only one $i=i_0$ for which $\alpha_{i_0}= \alpha^{'}_{i_0}\ne0$. Furthermore, we require that $\alpha_{i_0}=1$. 

So, in order to have a fusion of the type \eqref{fusgen}, we have two constraints:
\begin{enumerate}
\item $[g] \cdot [h]=[k]=[h]\cdot [g]$
\item $\exists! ~ \pi^{\omega}_k$ such that $m(\pi^{\omega}_k |_{N_{g} \cap N_{h} \cap N_k},\pi^{\omega}_{g}|_{N_{g} \cap N_{h} \cap N_k} \otimes \pi^{\omega}_{h}|_{N_{g} \cap N_{h} \cap N_k} \otimes \pi^{\omega}_{(g,h,k)})=1$ 
\end{enumerate}
The first constraint is on the conjugacy classes involved, and the second one is on the representations. The AH conjecture immediately implies that (1) is impossible for finite simple groups. Therefore, we see that
\begin{equation}\nonumber
{\rm AH\ conjecture}\Rightarrow{\rm no\ fusions\ as\ in\ \eqref{fusgen}\ for\ simple}\ G~.
\end{equation}
In particular, as claimed in the introduction, we see that
\begin{equation}\label{fusionLii}
\CL_{([g],\pi^{\omega}_g)}\times \CL_{([h],\pi^{\omega}_h)}\ne\CL_{([k],\pi^{\omega}_{k})}~,
\end{equation}
where $\CL_{([g],\pi^{\omega}_g)}=([g], \pi^{\omega}_g)$,  $\CL_{([h],\pi^{\omega}_h)}=([h], \pi^{\omega}_h)$, and $\CL_{([k],\pi^{\omega}_{k})}=([k], \pi^{\omega}_{k})$. So, in that language
\begin{equation}\nonumber
{\rm AH\ conjecture}\Rightarrow{\rm Theorem\ 1}~.
\end{equation}
Of course, this does not prove theorem 1 since the AH conjecture has not been proven. However, it is a non-trivial consistency check of the AH conjecture. We will prove theorem 1 in the next section.

Next, let us show how theorem 3 implies theorem 2. To understand this point, let us specialize the general fusion in \eqref{genfusion} to the product of a non-abelian Wilson line, $\CW_{\pi_1}=([1],\pi_1)$, with a non-abelian flux line, $\mu_{[h]}=([h],1_h^{\epsilon})$. In order to have such a flux line in our theory we should, as discussed in section \ref{fusionP}, either consider an untwisted discrete gauge theory or else a theory in which $\omega$ is such that $\eta_h\in H^2(N_h,U(1))$ is cohomologically trivial. 

To that end, we find
\begin{eqnarray}
N_{([1],\pi_1),([h], 1_h^{\epsilon})}^{([h],\pi^{\omega}_h)}&=&\sum_{(t,s)\in G \backslash G /N_h} m(\pi^{\omega}_h, {}^t\pi_{1}|_{N_h} \otimes {}^s1_{h}^{\epsilon} \cr&& \hspace{0.15cm}\otimes ~ \pi^{\omega}_{(1,h,h)}|_{N_h})~.
\end{eqnarray}
In this case, the double coset $G \backslash G/N_h$ is trivial. Hence, we have
\be
N_{([1],\pi_1),([h], 1_h^{\epsilon})}^{([h],\pi^{\omega}_h)}= m(\pi_h, \pi_{1}|_{N_h} \otimes 1_{h}^{\epsilon}\otimes \pi^{\omega}_{(1,h,h)}|_{N_h} )~.
\ee
In fact, the representation $\pi^{\omega}_{(1,h,h)}$ is trivial (this follows from the fixed nature of the $V_{1h}^h$ fusion space in the $G$-SPT \cite{Barkeshli:2014cna}). So the product of representations $\pi_{1}|_{N_h} \otimes 1_{h}^{\epsilon}\otimes \pi^{\omega}_{(1,h,h)}|_{N_h} $ is isomorphic to $\pi_{1} |_{N_h}\otimes 1_{h}^{\epsilon}$. Therefore, the expression above simplifies to 
\be\label{Nmrel}
N_{([1],\pi_1),([h], 1_h^{\epsilon})}^{([h],\pi^{\omega}_h)}= m(\pi^{\omega}_h, \pi_{1}|_{N_h} \otimes 1_h^{\epsilon})~.
\ee
Note that $\pi_1$ is an irreducible representation of $G$. Its restriction to $N_h$ is in general reducible. So $m(\pi_h, \pi_1|_{N_h}\otimes 1_h^{\epsilon})$ gives the multiplicity of the irreducible representation, $\pi_h$, in the decomposition of the representation, $\pi_1|_{N_h}\otimes 1_h^{\epsilon}$, into irreducible representations of $N_h$. If $\pi_{1}|_{N_h}$ is irreducible, $m(\pi_h, \pi_1|_{N_h}\otimes 1_h^{\epsilon})= \delta_{\pi_h, \pi_{1}|_{N_h}\otimes 1_h^{\epsilon}}$. Hence, we have the following fusion rules
\be
([1], \pi_1) \otimes ([h], 1_h)= ([h], \pi_1|_{N_h}\otimes 1_h^{\epsilon})~,
\ee 
if and only if $\pi_1|_{N_h}$ is an irreducible representation of $N_h$.

As a result, theorem 3 implies that we have more than one channel in the fusion
\begin{equation}\label{abcWmu2}
\CW_{\pi_1}\times\mu_{[h]}=\CL_{([h],\pi_h^{\omega})}+\cdots~.
\end{equation}
In fact, we may take the flux, $([h],1_h^{\epsilon})$, and replace it with a dyon, $([h], \pi^{\omega}_h)$. Note that, in some theories, such a dyon may exist while the flux line does not. We then find that the right-hand side of \eqref{Nmrel} becomes $m(\tilde\pi^{\omega}_h,\pi_1|_{N_h}\otimes \pi^{\omega}_h)$. Clearly, if the fusion in \eqref{abcWmu2} requires more terms on the right-hand side, so too will the fusion with the dyon replacing the flux. This is the content of theorem 2. 

Similarly, by the logic of this section, if we satisfy theorem 2 for the untwisted discrete $G$ gauge theory, we then have that, for any irreducible linear representation, $\pi_1$, of $G$ having dimension greater than one, $\pi_1|_{N_h}$ is reducible. This is the content of theorem 3. In conclusion, we have
\begin{equation}\nonumber
{\rm Theorem\ 3}\Leftrightarrow{\rm Theorem\ 2}~.
\end{equation}

Let us also note that we have chosen $\pi_1$ to be an irreducible representation of $G$ with dimension $>1$ so that $([1],\pi_1)$ is non-abelian. Hence, for the above fusion rule to be consistent, $\pi_{1}|_{N_h}$ should be an irreducible representation of $N_h$ of the same dimension.

What remains is to prove at least one of theorems 2 or 3. In the next section we give independent proofs of theorems 2 and 3. Proceeding through theorem 2 first gives us a more TQFT-flavored proof. Proceeding through theorem 3 first gives us a more group theory-flavored proof. We then conclude the next section by proving theorem 1 as well.

\section{Proofs of the cousin theorems}\label{proofs}
From the discussion in the previous section, to prove theorems 2 and 3 we need only prove one of them. However, each route has its own merits, so we give independent proofs of each. We follow by proving theorem 1 (which is logically independent of the others).

Let us first prove theorem 2. To that end, suppose we have a fusion of the form given in \eqref{abcWmu}, which we reproduce below for ease of reference
\begin{equation}\label{abcWmu3}
\CW_{\pi}\times\mu_{[g]}=\CL_{([h],\pi_h^{\omega})}~, \ \ \ g\ne1~.
\end{equation}
In section \ref{conjecture}, we argued that, if such a fusion exists, the electric charge of the dyon on the right-hand side is given by a reduction of an irreducible representation of the gauge group $G$ (i.e., $\pi_h^{\omega}=\pi|_{N_g}\otimes 1_{h}^{\epsilon}$) and $h=g$. Next, we note that the $S$-matrix satisfies \cite{Kitaev:2006lla}
\begin{equation}
S_{\CW_{\pi} \bar{\mu}_{[g^{-1}]}}={1\over|G|}{\theta_{\CL_{([g],\pi_g^{\omega})}}\over\theta_{\CW_{\pi}}\theta_{\mu_{[g]}}}d_{\CL_{([g],\pi_g)}}={1\over|G|}{\theta_{\CL_{([g],\pi_g^{\omega})}}\over\theta_{\CW_{\pi}}\theta_{\mu_{[g]}}}d_{\CW_{\pi}}d_{\mu_{[g]}}~,
\end{equation}
where $\bar\mu_{[g^{-1}]}$ is the conjugate of $\mu_{[g]}$. Therefore,
\begin{equation}\label{Smatrel}
|S_{\CW_{\pi} \mu_{[g]}}|={1\over|G|}d_{\CW_{\pi}}d_{\mu_{[g]}}~.
\end{equation}
Using \eqref{charconst}, we know that \eqref{Smatrel} implies $|\chi_{\pi}(g)|={\rm deg}\ \chi_{\pi}$, where $\chi_{\pi}$ is the character corresponding to the Wilson line's charge, and ${\rm deg}\ \chi_{\pi}=|\pi|>1$ is the dimension of $\pi$.

A standard argument in representation theory then implies that $\pi(g)=c\cdot\mathds{1}_{|\pi|}$, where $\mathds{1}_{|\pi|}$ is the $|\pi|\times|\pi|$ unit matrix, and $c$ is an $n^{\rm th}$ root of unity (the twist of the dyon). Next, choose some $k\in [G,g]:=\left<\ell g\ell^{-1}g^{-1}|\ell\in G\right>$. Clearly,
\begin{eqnarray}
\pi(k)&=&\pi(\ell g\ell^{-1}g^{-1})=\pi(\ell)\cdot c\cdot\mathds{1}_{|\pi|}\cdot\pi(\ell)^{-1}\cdot c^{-1}\cdot\mathds{1}_{|\pi|}\cr&=&\mathds{1}_{|\pi|}~.
\end{eqnarray}
Since $G$ is a simple group, we can choose $k\ne1$. As a result, $\pi$ is an unfaithful representation of $G$. Therefore, the kernel, ${\rm ker}(\pi)$, is a non trivial normal subgroup. Since $G$ is simple, we must have ${\rm ker}(\pi)=G$. But then, $\pi$ cannot be an irreducible representation. Note that we may repeat this proof verbatim by taking $\CL_{([g],\pi_g^{\omega})}$ instead of the flux line. Therefore fusion of the form in \eqref{abcWd} is also forbidden. $\square$ 

By the discussion in section \ref{conjecture}, we have also proved theorem 3. Although this proof is mathematical, it also has a distinctly TQFT-flavor: notice the prominent role of the modular $S$ matrix (and also, to a lesser extent, the twists).

Alternatively, we may also give a direct group theoretical proof of theorem 3 (and therefore of theorem 2 via section \ref{conjecture}) as follows:

Since $G$ is a non-abelian simple group, its irreducible representations of dimension larger than one must be faithful (otherwise their kernels would be non-trivial normal subgroups). Now, consider some faithful non-abelian representation, $\pi$. Furthermore, take some $g\in G$ such that $g\ne1$ and consider the centralizer, $N_g$.

Suppose the restriction $\pi|_{N_g}$ is irreducible. Clearly $g$ is central in $N_g$. As a result, by Schur's lemma
\begin{equation}
\pi|_{N_g}(g)=c\cdot\mathds{1}_{|\pi|}~,
\end{equation}
where $c$ is an $n^{\rm th}$ root of unity. Since this is a restriction of a representation of $G$, we must also have in the parent group that
\begin{equation}
\pi(g)=c\cdot\mathds{1}_{|\pi|}~,
\end{equation}
and so it follows that
\begin{equation}\label{kereq}
\pi(hgh^{-1}g^{-1})=\mathds{1}_{|\pi|}~.
\end{equation}
Since the group is simple, $g\ne1$ cannot be in the (trivial) center of $G$. As a result, there exists $h$ such that $hgh^{-1}g^{-1}\ne1$. The result in \eqref{kereq} contradicts the fact that $\pi$ is faithful. $\square$

Let us now prove theorem 1. We reproduce the forbidden \eqref{fusionL} for ease of reference
\begin{equation}\label{fusionL2}
\CL_{([g],\pi^{\omega}_g)}\times \CL_{([h],\pi_h^{\omega})}=\CL_{([k],\pi_{k}^{\omega})}~,\ \ \ g,h\ne1~,
\end{equation}
where, according to the discussion in the previous section, $[k]=[gh]$. Similarly to the case of theorem 2, we have that 
\begin{eqnarray}
S_{\CL_{([g],\pi^{\omega}_g)}\CL_{([h^{-1}],(\pi_h^{\omega})^*)}}&=&{1\over|G|}{\theta_{\CL_{([gh],\pi_{gh}^{\omega})}}\over\theta_{\CL_{([g],\pi^{\omega}_g)}}\theta_{\CL_{([h],\pi^{\omega}_h)}}}d_{\CL_{([gh],\pi_{gh}^{\omega})}}\cr&=&{1\over|G|}{\theta_{\CL_{([gh],\pi_{gh}^{\omega})}}\over\theta_{\CL_{([g],\pi^{\omega}_g)}}\theta_{\CL_{([h],\pi^{\omega}_h)}}}\cdot \cr&&\cdot d_{\CL_{([g],\pi_{g}^{\omega})}} d_{\CL_{([h],\pi_{h}^{\omega})}}~,
\end{eqnarray}
where $\CL_{([h^{-1}],(\pi_h^{\omega})^*)}$ is the conjugate of $\CL_{([h],\pi_h^{\omega})}$. Therefore,
\begin{equation}
|S_{\CL_{([g],\pi^{\omega}_g)}\CL_{([h],\pi_h^{\omega})}}|={1\over|G|} d_{\CL_{([g],\pi_{g}^{\omega})}} d_{\CL_{([h],\pi_{h}^{\omega})}}~.
\end{equation}
This last result allows us, as in the case of theorem 2, to use \eqref{charconst}. We then conclude that for any $\ell\in [g]$ and  $m\in [h]$, $\ell m=m\ell$ (i.e., that the two conjugacy classes $[h]$ and $[g]$ commute element-by-element).

Now, consider the product of conjugacy classes
\begin{equation}
[g]\cdot [g]=\sum_{[a]}N_{[g][g]}^{[a]}[a]~, \ \ \ N_{[g][g]}^{[a]}\in\mathbb{Z}_{\ge0}~.
\end{equation}
Clearly, we have that all elements on the left hand side commute with all elements of $[h]$. Therefore, the same is true of all elements in the conjugacy classes $[a]$. Now, consider taking pairwise products of all the $[a]$'s with themselves and with $[g]$ and so on. Eventually, we will come to a set of conjugacy classes closed under multiplication. This defines a normal subgroup $K\trianglelefteq G$ in which each element commutes with $[h]$. Since $G$ is simple, we must have that $K=G$. However, this means that $[h]$ commutes with all elements of the group and so we have a non-trivial center. This is a contradiction. $\square$

\section{Conclusion}
We have argued that $2+1$-dimensional discrete gauge theory is useful for putting conjectures and ideas involving finite simple groups into a broader context and unifying various relevant objects. Using this approach, we proved three theorems that TQFT relates to the AH conjecture.

In fact, we may also generalize the discussion in section \ref{conjecture} and show that the AH conjecture implies that, for any twisted or untwisted discrete gauge theory based on a non-abelian finite simple group, fusions of the form
\begin{equation}\label{fusionLfinal}
\CL_{([g],\pi^{\omega}_g)}\times \CL_{([h],\pi_h^{\omega})}=\sum_{\pi_{gh}^{\omega}} \CL_{([gh],\pi_{gh}^{\omega})}~,\ \ \ g,h\ne1~,
\end{equation}
are not allowed.

Finally, we argued that the lack of electric-magnetic dualities involving discrete gauge theories with non-abelian finite simple groups is a consistency check of our picture above and of the AH conjecture itself.

One natural question that remains is to better understand to what extent ideas involving non-abelian anyons can be used to prove the AH conjecture (see \cite{moori2011products,guralnick2019conjugacy,beltran2019new} and references therein for interesting recent progress on the AH conjecture). Since discrete gauge theories feature in various physical systems, perhaps we can hope for a physics proof of this conjecture. 

Theorem 3 is an example of the irreducible restriction problem for simple groups, in the special case of restriction of irreducible representations to centralizers. It will be interesting to explore its relationship with the Aschbacher-Scott program \cite{aschscott, kleschevtiep}.

Another interesting question is to understand to what degree fusion rules of the types we have been discussing constrain global properties of more general TQFTs. We will report on progress toward understanding this last question in \cite{toAppear}.

\acknowledgements{We are grateful for illuminating correspondence and discussions with D.~Aasen, M.~Fayers, S.~Ramgoolam, and I.~Runkel. M.~B. is funded by the Royal Society under the grant, \lq\lq New Constraints and Phenomena in Quantum Field Theory." M.~B. and R.~R. are funded by the Royal Society grant, \lq\lq New Aspects of Conformal and Topological Field Theories Across Dimensions." Our work is also partially supported by the STFC under the grant, \lq\lq String Theory, Gauge Theory and Duality."}

\newpage
\appendix{{\bf Appendix A: Direct proof of theorems 2 and 3 for \lq\lq AC" / \lq\lq CA" groups}}

Although we have given full proofs of theorems 3 and 2, it is amusing to give direct proofs that apply to certain classes of finite simple groups. For example, there is a large class of groups called \lq\lq AC" groups or, depending on the literature, \lq\lq CA" groups. These groups are defined to have abelian centralizers for all conjugacy classes of elements $g\ne1$. In this case, theorem 3 is trivially true: $\pi_1|_{N_g}$ is automatically reducible since $|\pi_1|>1$.

In particular, the $PSL(2,2^n)$ groups with $n\ge2$ are simple AC groups. In fact, these are the only such groups \cite{suzuki}. For $n=2$, we have $PSL(2,4)\simeq A_5$. More generally, however, the $PSL(2,2^n)$ groups are a distinct class of groups.

As a result, we conclude that in all (twisted or untwisted) discrete gauge theories based on AC groups our theorems hold. $\square$

\appendix{{\bf Appendix B: Direct proof of theorems 2 and 3 for $A_n$ groups}}

In this subsection, we will give special proofs of our theorems 2 and 3 for the case of  $A_n$ groups. In order for $A_n$ to be simple, we require $n=3$ or $n\ge5$ (proofs of the AH conjecture exist in the cases discussed here as well \cite{fisman1987proof}). The basic idea is to use Saxl's classification of irreducible characters of $A_n$, $\chi^{\lambda}$,  that remain irreducible upon reduction to a subgroup $G< A_n$ \cite{saxl1987complex}. We will argue that such subgroups cannot act as centralizers.

To that end, theorems 1 and 2 of \cite{saxl1987complex} constrain $\lambda$ and $G$ to be one of the following (note that $\Omega=\left\{1,2,\cdots, n\right\}$ is the set of elements $A_n$ acts on):
\begin{enumerate}
\item{$\lambda=(n)$ is the trivial representation}.
\item{$\lambda=(n-1,1)$ is the $n-1$ dimensional representation, and $G$ acts 2-transitively on $\Omega$.}
\item{$\lambda=(n-2,2)$, $n=9,11,12,23,24$ and $G$ is ${\rm P\Gamma L}(2,8)$, $M_{11}$, $M_{12}$, $M_{23}$, $M_{24}$ respectively.}
\item{$\lambda=(n-2,1,1)=(n-2,1^2)$ and either $n=2^c$ for some integer, $c$, with $G={\rm AGL}(c,2)$ or $n=11, 12, 12, 16, 22, 23, 24$ with $G=M_{11}, M_{11}, M_{12}, V_{16}A_7, M_{22}, M_{23}, M_{24}$ respectively.}
\item{$\lambda=(21,2,1)$ or $\lambda=(21,1^3)$, $n=24$, and $G=M_{24}$.}
\item{$\lambda=(\lambda_1^a)$ with $a\ne\lambda_1$, $n=a\lambda_1$, and $G=A_{n-1}$ stabilizing a point in $\Omega$.}
\item{$\lambda=(a^a)$, $n=a^2$, $G\ge A_{n-2}$, the stabilizer of two points in $\Omega$.}
\item{$\lambda=(a^b,b^{a-b})$, $n=(2a-b)b$, and $G=A_{n-1}$, the stabilizer of a point in $\Omega$.}
\item{$\lambda=(3^3)$, $n=9$, and $G={\rm P\Gamma L}(2,8)$ or $G={\rm AGL}(3,2)$.}
\item{$\lambda=(3^2,2)$, $n=8$, and $G={\rm AGL}(3,2)$.}
\end{enumerate}
Here we have used partitions of $n$ to label representations of $A_n$. 

In case (1) there is nothing to prove as the Wilson line would be the trivial abelian line. For case (2), the fact that $G$ is 2-transitive on $\Omega$ rules it out as a centralizer. To understand this statement, consider non-trivial $\sigma\in A_n$ and $g\in G$. Without loss of generality, we may take $\sigma(1)=2$. Since $G$ is 2-transitive on $\Omega$, it is also transitive, and we can choose $g$ so that $g(1)=3$. Now, $\sigma(3)=a\ne2$. By 2-transitivity, we may further choose $g$ such that $g(2)=2\ne a$. As a result $g^{-1}\sigma g(1)=g^{-1}\sigma(3)=g^{-1}(a)
\ne2=\sigma(1)$ and so $G$ does not centralize $\sigma$.

In case (3) we may check that there is no conjugacy class of length $|A_{9}|/|P\Gamma L(2,8)|$, $|A_{11}|/|M_{11}|$, $|A_{12}|/|M_{12}|$, $|A_{23}|/|M_{23}|$, or $|A_{24}|/|M_{24}|$ respectively. 

We may rule out the possibility of $G={\rm AGL}(c,2)$ in (4) by noting, as in \cite{neumann1975generosity}, that ${\rm AGL}(c,2)$ acts 2-transitively on the $c$-dimensional vector space over $GF(2)\simeq\mathbb{Z}_2$ \footnote{In fact, ${\rm AGL}(c,2)$ acts generously 3-transitively on the $c$-dimensional vector space over $\mathbb{Z}_2$ \cite{neumann1975generosity}.}. Since this vector space is $2^c$ dimensional, we may associate vectors with points in $\Omega$, and we are done by the logic that solved case (2). Since $G=V_{16}A_7$ acts 2-transitively on $\Omega$ \cite{saxl1987complex}, we see this will not work either. We may rule out the remaining possibilities in case (4) by similar logic to that employed in case (3). This logic also rules out case (5).

Let us now consider case (6). Here we may use the fact that non-trivial conjugacy classes in $S_n$ have length at least $n(n-1)/2$. As is well known, conjugacy classes in $A_n$ either have the same length as those in $S_n$ or else they have half the length. As a result, we conclude that non-trivial conjugacy classes in $A_n$ have length at least $n(n-1)/4$. This reasoning implies that $A_{n-1}$ is too large to act as a centralizer in $A_n$ (this statement holds since we can use GAP \cite{GAP} to explicitly check all cases $n\le11$; therefore, we need only worry about the cases $n>11$). This logic also rules out case (8). Cases (9) and (10) may be ruled out by explicit computation in GAP.

This leaves case (7). Here we may use the fact that $A_{n-2}$ fixes two points in $\Omega$ and acts $(n-4)$-transitively on the remaining $n-2$ points in $\Omega'\subset\Omega$. In fact, since we can check with GAP that this scenario doesn't arise for $n\le11$, we only need to discuss the case $n>11$ and use the weaker condition that $(n-4)$-transitivity implies 2-transitivity. Without loss of generality, we can again take non-trivial $\sigma\in A_n$ satisfying $\sigma(1)=2$. Without further loss of generality, there are three sub-cases to consider:
\begin{enumerate}[(a)] 
\item{$\sigma(2)=1$ and $\sigma(3)=4$.}
\vspace{0.03cm}
\item{$\sigma(2)=3$ and $\sigma(3)=1$.}
\vspace{-0.4cm}
\item{$\sigma(2)=3$ and $\sigma(3)=4$.}
\end{enumerate}

Let us consider $(a)$ first.  Suppose that $1,2\in\Omega'$. By transitivity, we can choose $g\in A_{n-2}$ such that $g(1)=x\ne1,2$ and $x\in\Omega'$. We then have $\sigma(x)=a\ne2$. By 2-transitivity, we may choose $g(2)=2\ne a$, and we have $g^{-1}\sigma g(1)=g^{-1}\sigma(x)=g^{-1}(a)\ne2=\sigma(1)$. Next, suppose that $1\in\Omega'$ but $2\not\in\Omega'$. Here we are forced to choose $g(2)=2$, but this doesn't matter. Indeed, the same logic we used when both $1,2\in\Omega'$ now works in this case as well. Continuing on, suppose instead that $1\not\in\Omega'$ but $2\in\Omega'$. Here we are forced to take $g(1)=1$. Since $2\in\Omega'$, we are free to choose $g(2)=y\ne2$. As a result, we have that $g^{-1}\sigma g(1)=g^{-1}\sigma(1)=g^{-1}(2)\ne 2=\sigma(1)$. To finish our discussion of $(a)$, let us suppose that $1,2\not\in\Omega'$. Then, $g(1)=1$ and $g(2)=2$. So $g^{-1}\sigma g(1)=2=\sigma(1)$. However, we have that $3,4\in\Omega'$. As a result, we can repeat our logic for the case $1,2\in\Omega'$ with $1,2\to3,4$.

Next consider $(b)$. This case can be treated identically to $(a)$ except for the scenario in which $1,2\not\in\Omega'$. However, the treatment here is similar. We must have $3\in\Omega'$ and so we can take $g(3)=x\ne1,2,3$. We also have $\sigma(x)=a\ne1$. Therefore, $g^{-1}\sigma g(3)=g^{-1}\sigma(x)=g^{-1}(a)\ne1$ (if $a\in\Omega'$, then this is clear since $1\not\in\Omega'$; if $a=2$, then $g^{-1}(a)=2$).

Finally, consider $(c)$. This case may be treated analogously to $(a)$. $\square$

\end{document}